\documentclass[acmtog]{acmart}
\usepackage{booktabs} % For formal tables

\AtBeginDocument{%
  \providecommand\BibTeX{{%
    \normalfont B\kern-0.5em{\scshape i\kern-0.25em b}\kern-0.8em\TeX}}}

\usepackage{graphics}

\usepackage[ruled]{algorithm2e} % For algorithms
\usepackage{soul}

\SetAlFnt{\small}
\SetAlCapFnt{\small}
\SetAlCapNameFnt{\small}
\SetAlCapHSkip{0pt}

\setcopyright{acmcopyright}\acmJournal{TOG}
\acmYear{2022}\acmVolume{41}\acmNumber{4}\acmArticle{137}\acmMonth{7} \acmDOI{10.1145/3528223.3530090}
\begin{document}
% Title portion
% \title{Real-time Transition Using Conditional Motion Manifold}
\title{Real-time Controllable Motion Transition for Characters}
% DO NOT ENTER AUTHOR INFORMATION FOR ANONYMOUS TECHNICAL PAPER SUBMISSIONS TO SIGGRAPH 2019!
\author{Xiangjun Tang}
\orcid{0000-0001-7441-0086}
\affiliation{%
 \institution{State Key Lab of CAD\&CG, Zhejiang University; ZJU-Tencent Game and Intelligent Graphics Innovation Technology Joint Lab}
 \country{China}}
\email{fcsx1tf@163.com}

\author{He Wang}
\orcid{0000-0002-2281-5679}
\affiliation{%
 \institution{University of Leeds}
 \country{United Kingdom}}
\email{H.E.Wang@leeds.ac.uk}

\author{Bo Hu}
\orcid{0000-0002-6599-7249}
\affiliation{%
 \institution{Tencent Technology (Shenzhen) Co., Ltd.}
 \country{China}}
\email{corehu@tencent.com}

\author{Xu Gong}
\orcid{0000-0003-3900-2903}
\affiliation{%
 \institution{Tencent Technology (Shenzhen) Co., Ltd.}
 \country{China}}
\email{xugong@tencent.com}

\author{Ruifan Yi}
\orcid{0000-0003-1693-6877}
\affiliation{%
 \institution{Tencent Technology (Shenzhen) Co., Ltd.}
 \country{China}}
\email{ryanfyi@tencent.com}

\author{Qilong Kou}
\orcid{0000-0003-0400-970X}
\affiliation{%
 \institution{Tencent Technology (Shenzhen) Co., Ltd.}
 \country{China}}
\email{rambokou@tencent.com}

\author{Xiaogang Jin*}
\thanks{*Corresponding author}
\orcid{0000-0001-7339-2920}
\affiliation{%
 \institution{State Key Lab of CAD\&CG, Zhejiang University; ZJU-Tencent Game and Intelligent Graphics Innovation Technology Joint Lab}
 \country{China}}
\email{jin@cad.zju.edu.cn}
%\affiliation{%
% \institution{Rajiv Gandhi University}
%\email{aprna_patel@rguhs.ac.in}
%\affiliation{%
%  \institution{Tsinghua University}
%  \streetaddress{30 Shuangqing Rd}
%  \country{China}
%}
%\email{chan0345@tsinghua.edu.cn}
%\author{Ting Yan}
%\affiliation{%
%  \institution{Eaton Innovation Center}
%  \city{Prague}
%  \country{Czech Republic}}
%\email{yanting02@gmail.com}
%\author{Tian He}
%\affiliation{%
%  \institution{University of Virginia}
%  \department{School of Engineering}
%  \city{Charlottesville}
%  \state{VA}
%  \postcode{22903}
%  \country{USA}
%}
%\affiliation{%
%  \institution{University of Minnesota}
%\email{tinghe@uva.edu}
%\author{Chengdu Huang}
%\author{John A. Stankovic}
%\author{Tarek F. Abdelzaher}
%\affiliation{%
%  \institution{University of Virginia}
%  \department{School of Engineering}
%  \city{Charlottesville}
%  \state{VA}
%  \postcode{22903}
%  \country{USA}
%}

\renewcommand\shortauthors{Tang et al.}

\begin{abstract}

Real-time in-between motion generation is universally required in games and highly desirable in existing animation pipelines. Its core challenge lies in the need to satisfy three critical conditions simultaneously: \textit{quality}, \textit{controllability} and \textit{speed}, which renders any methods that need offline computation (or post-processing) or cannot incorporate (often unpredictable) user control undesirable. To this end, we propose a new real-time transition method to address the aforementioned challenges. Our approach consists of two key components: motion manifold and conditional transitioning. The former learns the important low-level motion features and their dynamics; while the latter synthesizes transitions conditioned on a target frame and the desired transition duration. We first learn a motion manifold that explicitly models the intrinsic transition stochasticity in human motions via a multi-modal mapping mechanism. Then, during generation, we design a transition model which is essentially a sampling strategy to sample from the learned manifold, based on the target frame and the aimed transition duration. We validate our method on different datasets in tasks where no post-processing or offline computation is allowed. Through exhaustive evaluation and comparison, we show that our method is able to generate \textit{high-quality} motions measured under multiple metrics. Our method is also \textit{robust} under various target frames (with extreme cases). 

% Finally, it is lightweight and can be implemented as a standalone and a plug-in that can be conveniently embedded into existing software.

%After observing the varying information across joints in this task, we carefully choose two different representations for upper and lower joints, to help improve motion quality. Besides, building the matching component meets two requirements. 

% The first is consistency, which means that the matching component continuously synthesizes a motion with the correct relation of hip velocity and pose change for any reasonable input,  whatever it is from the training set or the output of transition component.
% The second requirement is diversity, which ensures the output space of the matching component covers as many motions as possible to synthesize a proper sequence to infill any start and end frames. 

% \HW{Real-time generation of high-quality and controllable character animation is universally required in most modern games. However, it is still an open problem after decades' research. The core challenge lies in the need to satisfy three key conditions: \textit{quality}, \textit{controllability} and \textit{speed}, which essentially rules out methods that need offline computation, post-processing and cannot incorporate (often unpredictable) user control. In this paper, we propose a new method, XXXX, to simultaneously address the aforementioned challenges. XXXX consists of two key components: matching and transitioning. The former relates the root velocity to the pose......
% }
\end{abstract}





\begin{teaserfigure}
\centering
  \includegraphics[width=\textwidth]{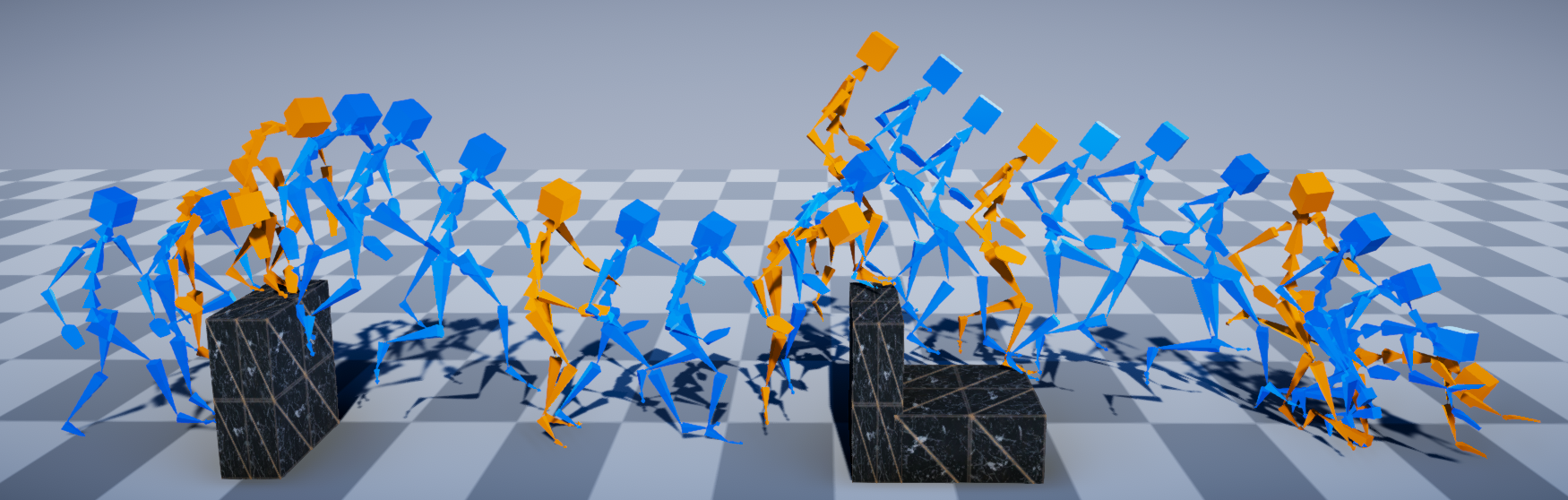}
  \caption{In-between motion sequences (blue) between target frames (orange) generated by our method. Given a target frame and a desired transition duration, the controlled character can dynamically adjust strategies, e.g., different step sizes, velocities, or motion types, to reach the target without visual artifacts. }
  \Description{}
  \label{fig:teaser}
\end{teaserfigure}

\maketitle
\newcommand{\HW}[1]{{\textcolor{red}{(#1)}}}

\section{Introduction}

In-between motion generation has been a long-standing problem in computer graphics/animation ~\cite{witkin1988spacetime}, and recently revived~\cite{zhang2018data,kaufmann_convolutional_2020,harvey2018recurrent} under the context of deep learning. It has been heavily relied upon in both offline animation pipelines and online motion synthesis in games. Speedy generation of high-quality motions without post-processing or offline computation is highly desirable in the former, and is often a must in the latter. 
% In-between motion generation has been a long-standing problem in computer graphics/animation ~\cite{rose1996efficient,witkin1988spacetime}. Recently, it has been revived~\cite{zhang2018data,kaufmann_convolutional_2020,harvey2018recurrent} under the context of deep learning. In-between motion has been heavily relied upon in traditional offline character animation pipeline and is also required for online motion synthesis in games. While speedy generation of high-quality motions without requiring post-processing or offline computation can greatly boost productivity in the former, it is often a must in the latter. 
% Therefore, we focus on this classic and fundamental problem in this paper.

Early methods formulate in-between motions as motion planning problem~\cite{Ye_synthesis_2010,wang_harmonic_2013,wang_energy_2015}, which requires solving complex optimizations and are prohibitively slow for real-time applications. Data-driven methods have also been developed~\cite{Kovar_motion_2008,Min_motion_2012,shen_posture_2017}. However, to handle arbitrary in-between motions and target frames, the size of needed data in memory grows exponentially~\cite{harvey_robust_2020}. In the era of deep learning, in-between motions can be interpreted as a motion manifold learning problem~\cite{Wang_STRNN_2019,Chen_dynamic_2020,holden_deep_2016}, or a control problem~\cite{ling_character_2020} if dense temporal control signs are available. Compared with previous data-driven methods, deep neural networks can leverage compressed data representations, but cannot be easily converted into in-between motion generators~\cite{harvey_robust_2020}. 
Very recently, this classic problem has been revived~\cite{duan_single-shot_2021,kaufmann_convolutional_2020},  but there is still a lack of model generality when facing arbitrary target frames, which is often the case especially in real-time games where the user input is unpredictable.

There are two major challenges in real-time in-between motion generation. The foremost is the \textit{motion quality}. Since motions need to be generated fast, post-processing is highly undesirable. Also, offline computation and any human intervention are strictly ruled out. One possible solution is a motion model which can capture the fine-grained dynamics of diverse actions and act as a source of motion generation. Designing such a model needs to consider the intrinsic transition ambiguity of human motions, i.e. multiple frames or actions could follow a given one. This leads to the second challenge: \textit{controllability}. While capturing and disambiguating the transitions can achieved by relying on continuous control signals~\cite{holden2017phase}, our problem setting only involves sparse target frames. The control sparcity differentiates our problem from those with similar key frames or dense control signals. In addition, the generated motion needs to satisfy the target frame and the aimed transition duration simultaneously. Failing in transition disambiguation will lead to `averaged' motions~\cite{fragkiadaki2015recurrent}, while failing in controllability will break the constraints imposed by the user.

In this paper, we propose a novel method which can generate high-quality in-between motions in real-time, given the starting and end frame with the desired period of transition. Our method consists of two components designed to address the aforementioned challenges. We start by representing the natural \textit{motion manifold} and focus on modeling the multi-modality of motion transitions under a Markov assumption. To incorporate the target frame and the desired transition period, we further propose a new \textit{sampler} which samples from the learned motion manifold, under the constraints imposed by the initial, target frame and the desired transition period.

% The motion manifold employs a Conditional Variational Autoencoder (CVAE) architecture. Our CVAE learns a conditional distribution of transitions between frames~\cite{ling_character_2020}, with an embedded multi-modal mapping to model the transition ambiguity, which is realized as a Conditional Mixture of Experts (CMoEs) in the latent space. As a result, our motion manifold can act as a high-quality source for online motion synthesis.
The motion manifold employs a Conditional Variational Autoencoder (CVAE) architecture. Instead of learning a conditioned latent distribution of original data as traditional CVAEs, our CVAE learns a conditional distribution of transitions between frames~\cite{ling_character_2020}. Further, we explicitly model the transition ambiguity as a multi-modal mapping between frames, by utilizing a Conditional Mixture of Experts (CMoEs) in the latent space and the decoding phase. As a result, our motion manifold can act as a high-quality representation which provides a reliable source for online motion synthesis. The other key component is a transition sampler which samples one frame at a time. The sampler is realized as a deep neural network which models the dynamics of the generated motion by a Recurrent Neural Network (RNN). It conditions the next frame on the current frame, the target frame, the desired transition period and the remaining motion, through a multi-step residual network architecture.

We test our method on two popular datasets, under a variety of conditions, e.g. action types, generation lengths, the spatio-temporal aspects of the target frame and transition period. We employ both qualitative and quantitative evaluation, with multiple metrics including reconstruction errors, foot skating, and Normalized Power Spectrum Similarity (NPSS). After exhaustive ablation studies among different alternative architectures and representations, and comparisons with the state-of-the-art methods, we show that our method can generate high-quality motions in real-time, is robust across action types and dynamics, generalizes well to extreme user inputs, and outperforms existing methods under multiple criteria.

Our main contributions can be summarized as follow:
\begin{itemize}
    \item We present a novel online framework for high-quality real-time in-between motion generation without post-processing.
    \item We propose a natural motion manifold model which is able to condition motion transitions on control variables for transition disambiguation, simultaneously providing controllability and ensuring motion quality.
    % We present a hip-conditional motion manifold that avoids sampling drifting motion and improves generalization in the wild.
    % We present a motion manifold that encodes the pose distribution under hip velocity so that the character always adopts a reasonable strategy for a different movement. \HW{This doesn't sound very novel. }
\end{itemize}

\section{Related work}
Existing methods formulate in-between motions in various fashions. In the early days, in-between motions were often formulated as a motion planning problem~\cite{wang_energy_2011,arikan_interactive_2002,Safonova07constructionand,beaudoin2008motion,levine2012continuous}, where fairly sophisticated motions can be synthesized. Complex optimization problems~\cite{chai_constraint-based_2007} are formed with respect to various constraints such as contact and control input, leading to slow computation which is impractical for the animators and impossible for real-time applications. Alternatively, data-driven methods can avoid slow optimizations by searching in structured data, e.g. motion graphs~\cite{Kovar_motion_2008,Min_motion_2012,shen_posture_2017}. However, since the control or constraints can be diverse, the size of needed data in memory to cover all situations grows exponentially~\cite{harvey_robust_2020}, leading to unaffordable space complexity. Recently, in-between motions have been interpreted as a motion manifold learning problem~\cite{Wang_STRNN_2019,Chen_dynamic_2020,holden_deep_2016,li2021task,rempe2021humor,petrovich_action-conditioned_2021}, or a control problem~\cite{ling_character_2020} in deep learning. Compared with previous data-driven methods, deep neural networks can leverage compressed data representation~\cite{holden_learned_2020}. However, they cannot be easily converted into an in-between motion generator~\cite{harvey_robust_2020}. 

One attempt to convert the deep neural network to an in-between motion generator is to add constraints as regularization in the loss function. Examples include RNN-based models~\cite{martinez2017human,chiu2019action} which can generate motions with constraints to reduce motion ambiguity~\cite{harvey2018recurrent}. However, simply adding constraints cannot achieve high-quality results when different transition duration is needed. Subsequently, a time-to-arrival condition is proposed~\cite{harvey_robust_2020} and a generative adversarial network (GAN) is employed to safe-guard the quality of the generated motion.
However, without explicitly extracting different hierarchies in human dynamics~\cite{chiu2019action} or utilizing the relation of joints~\cite{jain_structural-rnn_2016}, the diversity of generated sequence is limited, leading to poor generalizability to unseen constraints such as extreme user control.

If real-time performance is not a requirement, offline methods can be employed in motion completion such as in-between motion generation or joints filling. Motion completion can be solved by optimizing the sampling of the motion manifold~\cite{li2021task}, or
% \st{The completion can be}
considered as an analogy to the image infilling problem~\cite{kaufmann_convolutional_2020,hernandez2019human}. A convolutional network can be employed to infill the missing parts of the sequence. The missing parts do not have to be whole frames. They could be just the position or orientation of a single joint. {Besides,  separating local motions and global trajectories enables convolutional networks to focus on generating realistic local poses \cite{zhou2020generative}. } {The infilling problem can also be solved by Transformers~\cite{duan_single-shot_2021}. Instead of padding the missing frames, Transformer-based motion infiller~\cite{yuan2021glamr} can restrict its attention to visible frames to achieve effective temporal modeling.} Time efficiency is normally not the primary goal of offline methods, so it is acceptable to utilize post-processing to improve the motion quality. Nevertheless, we aim for real-time generation.

\begin{figure*}[]
	\centering
	\includegraphics[width=1\linewidth]{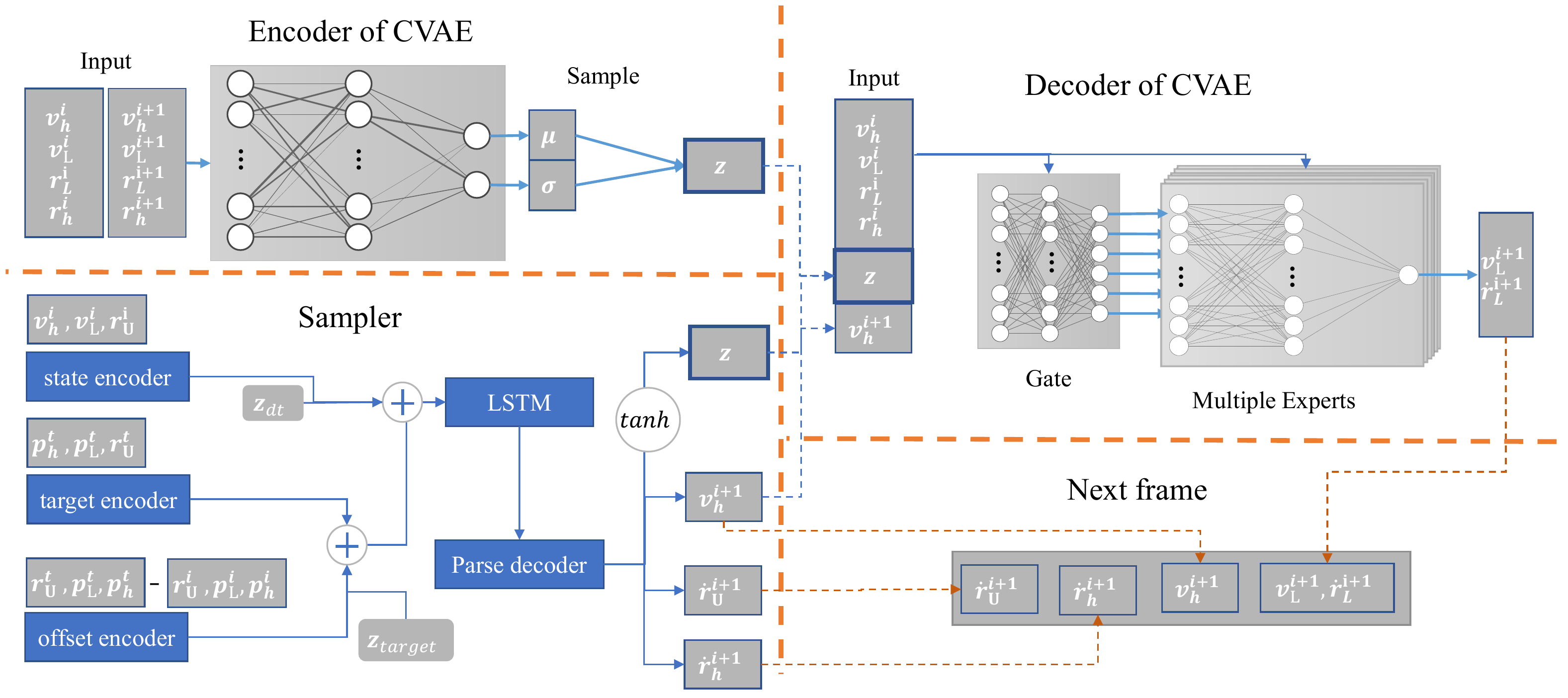}
    \caption{{When training the CVAE, the encoder takes frames $i$ and $i+1$ as input and generates the mean value $\mu$ and log variance $\sigma$ of the normal distribution.  The motion manifold model (the decoder of CVAE) takes the current frame $(v_h^{i}, \bold{v}_L^{i}, \bold{r}_L^{i}, r_h^{i})$, the character movement $v_h^{i+1}$ at the next frame,  and a latent vector $z$ to generate the rotation and velocity of the lower-body joints. When training the transition sampler, we first remove the encoder of the CVAE, then fix the decoder and connect the transition sampler to the fixed decoder to train the sampler. The target frame, current frame, and offset are encoded by the target, state, and offset encoders, respectively. Subsequently, the embeddings $z_{dt}$ and $z_{target}$ are added to the encoded vectors, and an LSTM network takes the encoded vectors to produce the next state. Finally, the parse decoder takes the state and outputs the upper joints $\dot{r}_U^{i+1}$ and the sample $(z,v_h^{i+1})$. }}
	\label{fig:encoder}
	\label{fig:decoder}
	\label{fig:transition}
\end{figure*}

\section{Methodology}

Our method consists of two main components: a natural motion manifold model and a sampler for motion generation. We first introduce the natural motion manifold that learns the low-level short-horizon motion dynamics. We then introduce a sampling strategy to generate motions from the learned manifold satisfying the target frame and the aimed transition duration. 
% The details of the implementation, training and corresponding parameters can be found in the supplemental material.

\subsection{The Motion Manifold}
\label{sec:motionManifold}
To generate a motion $M=\{S^1,\dots, S^{n-1}\}$ with $n-1$ frames, each frame is denoted by $S^i = \{\mathbf{p}^i_L, \mathbf{p}^i_h, \mathbf{p}^i_U, \mathbf{r}^i_L, \mathbf{r}^i_h, \mathbf{r}^i_U,\mathbf{v}^i_L, \mathbf{v}^i_h, \mathbf{v}^i_U\}$ where $\mathbf{p}$, $\mathbf{r}$ and $\mathbf{v}$ are the joint position, rotation and velocity, and the subscript $L$, $h$ and $U$ indicate the lower body, the hip and the upper body joints respectively. Given a starting frame $S^0$, a target frame $S^t$, and the aimed transition duration $z_{dt}$, the joint probability of $M$ can be represented as:
\begin{align}
\label{eq:jointProb}
P(M) =  \int\int\int P(M | S^0, S^t, z_{dt}) P(S^0, S^t, z_{dt}) dS^0 dS^t dz_{dt},
\end{align}
where we assume the independence among $S^0$, $S^t$, and $z_{dt}$ and omit them for later. Under a Markov assumption, $P(M)$ can be decomposed into:
\begin{equation}
    P(M) = \prod_{i=1}^{n-2} P(S^{i+1}|S^i),
\end{equation}
where $P(S^{i+1}|S^i)$ can be learned in many ways, e.g. through recurrent models~\cite{Wang_STRNN_2019} or co-embedding of consecutive frames~\cite{ling_character_2020}. Here, we choose to employ the co-embedding strategy as it easily allows conditional variables to be introduced. We introduce a latent variable $z$ to encode the co-embedding of two consecutive frames and also use the next frame hip velocity $v_h^{i+1}$ as a conditional variable. While $z$ encodes the transition probability of two consecutive frames (a.k.a the dynamics), $v_h^{i+1}$ can help disambiguate the next frame, for which we will give details when introducing the transition sampling. Introducing $z$ and $v_h^{i+1}$ into $P(S^{i+1}|S^i)$ gives:
\begin{equation}
    P(S^{i+1}|S^i) = \int P(S^{i+1}|S^t,z,v_h^{i+1})P(z)P(v_h^{i+1}),
\end{equation}
Note that we divide all joints into three groups: upper-body, hip, and lower-body. This is because we empirically find they have different importance in the generation. The hip velocity gives a strong indication of the next frame (e.g. distinguishing between motions with high and low velocities). The lower-body joints significantly influence the visual quality due to potential foot sliding. The upper-body joints are less constrained comparatively. Therefore, we focus on learning the lower-body and the hip in $P(S^{i+1}|S^i)$:
% \begin{align}
% \label{eq:cvae}
%     P(S^{i+1}|S^i) = \int P(v^{i+1}_L,\dot{r}^{i+1}_L|v^{i}_h,v^{i}_L,r^{i}_h,r^{i}_L,z,v_h^{i+1})P(z),
% \end{align}
{
\begin{align}
\label{eq:cvae}
    P(S^{i+1}|S^i) = \int P(v^{i+1}_L,\dot{r}^{i+1}_L|c_{h,L},v_h^{i+1},z)P(z),
\end{align}
where $c_{h,L}^i = \{v^{i}_h,v^{i}_L,r^{i}_h,r^{i}_L\}$ consists of the lower-body and the hip joints of the current frame. We assume that $v_h^{i+1}$ is given during prediction and hence its prior distribution can be removed.}
% where $v_h^{i+1}$ is assumed to be given during prediction and hence its prior distribution is removed.
$\dot{r}$ is the angular velocity. Since $z$ should encode two consecutive frames, it can be independently learned via 
% \st{$P(z) = P(z|v^{i}_h,v^{i}_L,r^{i}_h,r^{i}_L, v^{i+1}_h,v^{i+1}_L,r^{i+1}_h,r^{i+1}_L)$}
{$P(z)=P(z|c_{h,L}^i,c_{h,L}^{i+1})$ or expanding it to $P(z)=P(z|c_{h,L}^i,v^{i+1}_h,v^{i+1}_L,r^{i+1}_h,r^{i+1}_L)$}. If we assume $z\sim N(0, \mathbf{I})$, where $N$ is a normal distribution, {$P(z)$ can then be considered as the encoder of a Conditional VAE or CVAE where the condition is $\{c_{h,L}^i,v_h^{i+1}\}$} and the latent space distribution is constrained to be a Normal, as shown in Figure~\ref{fig:encoder}. 

% \begin{figure}[h]
% 	\centering
% 	\includegraphics[width=1\linewidth]{Image/MatchingEncoder.pdf}
%     \caption{The encoder takes frame $i$ and $i+1$ as input. It has two hidden layers with 256 units followed by ELU activation. The two output layers generate the mean value $\mu$ and log variance $\sigma$ of the normal distribution, respectively.}
% 	\label{fig:encoder}
% \end{figure}

Next, 
% \st{$P(v^{i+1}_L,\dot{r}^{i+1}_L|v^{i}_h,v^{i}_L,r^{i}_h,r^{i}_L,z,v_h^{i+1})$}
{$P(v^{i+1}_L,\dot{r}^{i+1}_L|c_{h,L}^{i},v^{i+1}_h,z)$ can be considered as the decoder of the CVAE. Instead of reconstructing $r^{i+1}_L$ directly, we compute it by $r_L^{i+1} = \dot{r}_L^{i+1}+r_L^i$.} 
% \st{Note our CVAE is different from standard ones where the encoder is also conditional. Our encoder is unconditioned, encoding all frame transitions in a dataset, and leaving the disambiguation job to the decoder, which effectively reduces the size of the model for faster training.}
The conditional variables of the decoder are specifically designed to capture the ambiguous transitions that are intrinsic to human motions. It aims to learn discriminative transitions via a multi-modal mapping. Specifically, given a non-discriminative embedding $z$, the decoding is conditioned on the current frame ($v^{i}_h,v^{i}_L,r^{i}_h,r^{i}_L$) and the future hip velocity ($v_h^{i+1}$). Such a decoding process requires the decoder to learn a multi-modal mapping that is similar to incorporating different dense control signals~\cite{zhang_mode-adaptive_2018}~\cite{holden2017phase}. Therefore, we employ a Conditioned Mixture of Experts (CMoEs) model in the decoder, as shown in Figure~\ref{fig:decoder}. During learning, the CMoEs can learn discriminative mappings where each expert network tends to focus on learning one phase of motions. We also add a gating network which learns a weighting scheme for experts given a specific input. The final output is a weighted sum of all expert outputs.

\subsubsection{Losses}
To train the CVAE, we minimize a loss function:
\begin{equation}
    L = L_{foot} + L_{bone} + L_{rec} + L_{kl},
\end{equation}
where several loss terms are proposed. $L_{foot}$ is a foot skating loss in the joint position space:
\begin{equation}
    L_{foot} = \hat{v}_{end}+v_h,
    \label{eq:foot}
\end{equation}
where $\hat{v}_{end}$ is the predicted relative velocity of the contacting foot with respect to the ground. When the velocity is less than $0.2cm/s$, we assume a foot contact with the ground.
Although a joint angle representation is also theoretically possible with forward kinematics, the relation to be learned would become unnecessarily non-linear.

Besides, we add a bone length loss. For each joint $j$ and its neighbor joints in $n(j)$, the loss is:
\begin{equation}
     L_{bone} = ||\hat{p}_j - \hat{p}_k||_2 - ||p_j - p_k||_2, \forall{k = n(j), j\in \bold{L}},
     \label{eq:bone}
\end{equation}
where $\hat{p}_j$ is the predicted position of joint $j$.
% However, because keeping the joint's velocity equal to zero suffice the requirement, the training might have a local minimum problem when adding the bone length loss, a point we further discuss next.

The reconstruction loss is defined as the mean squared error (MSE) between the predicted pose and the ground-truth:
\begin{equation}
    L_{rec} = ||\hat{\bold{p}}_L-\bold{p}_L ||_2^2 + ||\hat{\bold{r}}_L-\bold{r}_L||_2^2.
\end{equation}
Finally, a KL-divergence loss is employed to constrain the distribution of the latent vector to be a standard Gaussian distribution:
\begin{equation}
\label{eq:kl}
    L_{kl} = -0.5\cdot (1+\sigma-\mu^2-e^{\sigma}),
\end{equation}
where $\mu$ and $\sigma$ are the mean and log variances.

\subsection{Transition Sampling}
Although the CVAE can learn a natural manifold, it can only perform uncontrolled generation. This is because $S^0$ can be easily used to start the generation in Equation~\ref{eq:cvae}, but the distributions are not conditioned on $S^t$ and $z_{dt}$. Explicitly learning $P(M)$ conditioned on $S^t$ and $z_{dt}$ requires learning the reverse Markov chain across all possible duration, which is not trivial. Therefore, we use a neural network to learn them implicitly. To be able to generate motions continuously, we need to sample $z$ and $v_h^{i+1}$ to generate $S^{i+1}$ given $S^i$, under the constraints of $S^t$ and $z_{dt}$. So the network is essentially a sampler for sampling frames from the learned manifold.

The architecture of the network is shown in Figure~\ref{fig:transition}. The sampler considers the constraints by the target encoder and the offset encoder, which encode the target frame and the offset between the current and the target frame, respectively. The key output is the next-frame condition $z$ and $v_h^{i+1}$. In addition, when used for decoding the next frame, $z$ and $v_h^{i+1}$ will be pulled through the decoder of our CVAE, where essentially a manifold projection is conducted to refine the pose. We also add a time-varying noise $z_{target}$ to the encoded vector, sampled from a zero-centered Gaussian distribution with variance equal to $0.5$. Its amplitude $\lambda$ decreases as it approaches the target frame so that the sampler's attention only focuses on the target when close to it. It also helps to improve the robustness to new conditioning information~\cite{harvey_robust_2020}. The amplitude of the noise decreases by the function:
\begin{equation}
    \lambda = clamp \left(\frac{dt-t_{zero}}{t_{period}-t_{zero}},0,1 \right),
\end{equation}
where $dt$ is the frame difference between the current time and the target, $t_{zero}$ is the frame duration without noise, and $t_{period}$ is the period of linear decrease of the noise. We empirically set $t_{zero}=5$ and $t_{period}=30$ in our experiments.

The sampler takes the current state of the pose via the state encoder. The constraint of $z_{dt}$ is represented by the time embedding $\bold{z}_{dt}$~\cite{harvey_robust_2020},  added to the latent vector of all encoders. The time embedding vector is similar to the positional encoding in \cite{vaswani2017attention}:
\begin{equation}
    \bold{z}_{dt,2i} = sin \left(\frac{dt}{10000^{2i/d}}\right),\quad \bold{z}_{dt,2i+1}=cos \left(\frac{dt}{10000^{2i/d}}\right),
\end{equation}
where $d$ represents the dimension of $\bold{z}_{dt}$  and $i \in [0,...,d/2]$ represents the dimension index.

Next, the recurrent neural network takes all latent vectors to predict the next state. A decoder parses the state to generate the sample $(z,v_h^{i+1})$ and the upper joints $\dot{r}_U^{i+1}$.

When passing the frame into encoders, we represent $S$ by the hip velocity, the lower joints' velocity, and the upper joints' rotation to reduce dimensionality  $(v_h,v_L,r_U)$ compared to the full state, and calculate the offset using the lower joints' position $p_L$. To balance the attention on the lower joints and upper joints, we apply z-score normalization on $p_L$ before passing it into the offset encoder.

% The separation hugely eases the complexity of transition missions. 

% Firstly, the hip velocity pattern is relatively straightforward to be learned when given the target position and time horizons. Besides, the accuracy of upper joints is not emphasized because they hardly induce drifting and hardly be unnatural.  The core requirement for upper joints is smoothly interpolating to the target. In addition, the manifold ensures motion quality if the latent vector $z$ is sampled from a Gaussian distribution, which is also easy to be achieved. 

% \begin{figure}[]
% 	\centering
% 	\includegraphics[width=1\linewidth]{Image/TransitionAr.pdf}
%     \caption{The target frame, current frame, and offset are encoded by the target, state and offset encoder, respectively. Subsequently, the embeddings $z_{dt}$ and $z_{target}$ are added to the encoded vectors. Then an LSTM takes the encoded vectors to produce the next state. Finally, the parse decoder takes the state and outputs the upper joints $\dot{r}_U^{i+1}$ and the sample $(z,v_h^{i+1})$. \TXJ{The LSTM has 1024 units in the hidden layer. The parse decoder has two hidden feed-forward layers with 512 units in the first layer and 256 units in the second layer. Both layers are followed by ELU.}}
% 	\label{fig:transition}
% \end{figure}

All three encoders are two-layer feed-forward networks with 512 units in the first hidden layer and 256 units in the second layer. Each layer is followed by PLU activation. The parse decoder has three layers with 512 units in the first hidden layer and 256 units in the second layer, followed by ELU activation. To compute $z$, which is a key input to our CVAE to sample the next frame, we apply a tanh function and scale the output by 4.5 to ensure a good coverage of the normal distribution. 

\subsubsection{Losses}
To train the sampler, we propose the following loss function:
\begin{equation}
    L = L_{rot} + L_{leg} + L_{pos,rot},
\end{equation}
where $L_{rot}$ is a L1 norm rotation loss for all joints and $L_{leg}$ is a position loss for lower-body joints: 
\begin{equation}
    \begin{array}{ll}
         L_{rot} = ||\hat{\bold{r}}_L-\bold{r}_L ||_1 + ||\hat{\bold{r}}_U-\bold{r}_U ||_1,  \\
         L_{leg} = ||\hat{\bold{p}}_L-\bold{p}_L ||_1.
    \end{array}
\end{equation}
Besides, similar to Harvey \textit{et al}.~\shortcite{harvey_robust_2020}, we employ Forward Kinematics (FK) to obtain the position $\hat{\bold{p}}_{rot}$ of all joints from their predicted rotation. The loss between $\hat{\bold{p}}_{rot}$ and the ground truth position $\bold{p}$ helps to implicitly weigh the rotation of the bone’s hierarchy for better results~\shortcite{pavllo2019modeling}:

\begin{equation}
 L_{pos,rot} = ||\hat{\bold{p}}_{rot}-\bold{p} ||_1.
\end{equation}
In addition, the foot skating loss (see Eq.~\ref{eq:foot}) and the bone length loss (see Eq.~\ref{eq:bone}) are also used for training the sampler. 

\section{Implementation}
\begin{figure}[]
	\centering
	\includegraphics[width=1.0\linewidth]{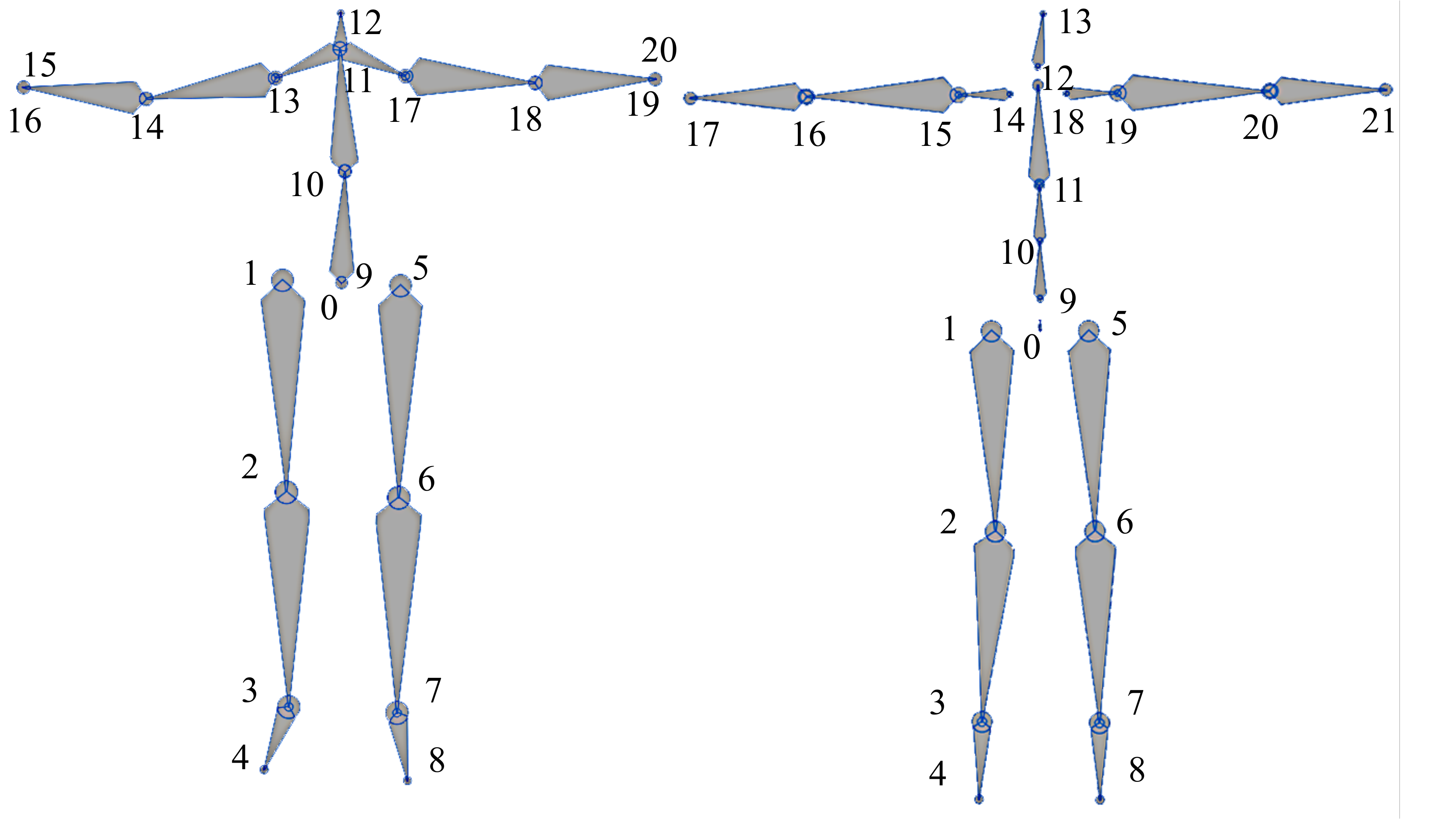}
    \caption{The left character has 21 joints from the Human3.6M. The right character from the Lafan1 dataset has 22 joints. In our setup, the lower joints set is $L = \{ 1, 2, 3, 4, 5, 6, 7, 8 \}$ , the joint $0$ is hip joint, and the other joints are the upper joints. }
	\label{fig:skeleton}
\end{figure}
\subsection{Data formatting}

We use the LaFAN1 dataset~\cite{harvey_robust_2020} and the Human3.6M dataset~\cite{IonescuSminchisescu11}~\cite{h36m_pami}.  We remove the wrist and thumb joints from the Human3.6M dataset, which leaves us with $21$ joints. The character from the Lafan1 dataset has $22$ joints. We employ different representations for different joints. As shown in Fig.~\ref{fig:skeleton}, we use the position-based representation for $8$ lower joints and the rotation-based representation for upper joints. All lower joints connect less than two other joints to determine their orientation.

Zhang et al.~\shortcite{zhang_mode-adaptive_2018} proposed to represent joint rotation by a 2-axis rotation matrix $r\in \mathbf{R}^6$, containing a 3D vector for the up direction and a 3D vector for the forward direction. The joint position $j$ contains a 3D vector $v_j \in \mathbf{R}^3$ to represent the velocity of the joint and a 3D vector to represent the up direction. We replace the 3D up direction with the 2-axis rotation matrix for uniformity. So the lower joints $X_L^i$ of frame $i$ can be represented by eight lower joints' rotation $\bold{r}_{L}^i \in \bold{R}^{8\times 6}$ and six lower joints' position $\bold{p}_{L}^i \in \bold{R}^{6\times 3}$ with the velocity $\bold{v}_{L}^i \in \bold{R}^{6\times 3}$. Notice that we discard the velocity of joints 1 and 5 because they are determined by joint 0's rotation. 

{Both datasets contain multiple subjects. The subjects in the test set are different from the training set in our experiments, which ensures that our motion model can generalize to different subjects after training.}
The Lafan1 dataset contains 496,672 motion frames performed by 5 motion subjects. Similar to \cite{harvey_robust_2020,duan_single-shot_2021}, we use subject 5 as the test set.
For the Human3.6M dataset, we choose to use a subset containing the walk-related actions (walking, walkingdog, walkingtogether), the same as~\cite{harvey_robust_2020} did. We work with a 25HZ sampling rate and take Subject 5 as the test subject.

{For both datasets,  we split the training set into multiple 50-frames windows. Similar to~\cite{harvey_robust_2020}~\cite{duan_single-shot_2021}, two consecutive windows have 25 overlapped frames.}

When training, the input of the current step is the output from the last step so that the error accumulates as the generated sequence grows. This effect increases the robustness because the network learns a pose not only from the dataset.

\subsection{Training of Motion Manifold}

The 50-frames sequence is still too long for learning motion manifold to convergence efficiency. We equally split the 50-frames sequence into two 25-frames sequences before training the architecture. 

The bone length loss constrains the joints' velocity to zero, generating a weird motion sequence. To avoid this, we train the architecture twice. We first use the reconstruction loss and KL-divergence loss, teaching the network to predict an approximate pose. Secondly, we add foot skating loss and bone length loss. 

{Inspired by Ling et al.~\shortcite{ling_character_2020}, we pass the latent variable $z$ and the future hip velocity $v_{h}^{i+1}$ to every layer of the expert network to avoid posterior collapse. The encoder of the CVAE has two hidden layers with 256 units followed by ELU activation. The gating network has two hidden feed-forward layers followed by ELU activation. The output layer of the gating network uses Softmax activation. Each of expert network is a three-layer feed-forward network with 256 units in the hidden layers followed by ELU. In our preliminary experiments, six expert networks achieve good results and fewer than six lead to worse accuracy. We therefore empirically set the expert number to six in our implementation.
}

The scheduled sampling strategy is employed the first time. The network takes the predicted pose the last timestep as input with probability $p$. Otherwise, it takes the ground truth from the dataset as input. The probability starts at 0 for the first $k$ epochs and then increases to 1 linearly for another $k$ epochs. We set $k=5$ for the Lafan1 dataset and $k=20$ for the Human3.6M dataset.

We use AMSgrad optimizer with adjusted parameters ($\beta_1=0.5, \beta_2=0.9$). 
At the first training time, the learning rate is initialized to 1e-4 and linear decreases to 1e-5 by 50,000 iterations. The learning rate starts at 0 for the second time and increases to 1e-4 for ten epochs so that the added losses do not significantly change the network's parameters. After ten epochs, the learning rate decreases with the same decreasing rate as the first time.
We scale all losses to be approximately equal to 1 for an untrained network without employing extra weights.

For the transition sampler, the target frame, current frame, and offset are encoded by the target, state and offset encoder, respectively. Subsequently, the embeddings $z_{dt}$ and $z_{target}$ are added to the encoded vectors. Then an LSTM takes the encoded vectors to produce the next state. Finally, the parse decoder takes the state and outputs the upper joints $\dot{r}_U^{i+1}$ and the sample $(z,v_h^{i+1})$. 

\subsection{Training of Transition Sampler.}

{After training the CVAE, we remove its encoder, fix its decoder, and connect the transition sampler to the fixed decoder to train the sampler.}

{All encoders are feed-forward networks with a hidden layer of 512 units and an output layer of 256 units. All layers use PLU as the activation. The LSTM has 1024 units in the hidden layer. The parse decoder has two hidden feed-forward layers with 512 units in the first layer and 256 units in the second layer. Both layers are followed by ELU.}

During training, we sample a transition length from 5 to 30 frames from a window in each learning step so that the network can learn from different transition lengths and the target frames.

The AMSgrad optimizer is also employed for training the transition architecture. The learning rate equals 1e-3, the weights for $L_{rot}, L_{leg}$ are $1$ and the weights for $L_{pos,rot}, L_{bone}, L_{foot}$ equals 0.5. We train the transition architecture for 300,000 iterations costing approximately one day.

\section{Experiments and Results}
\label{sec:experiments}

We conduct our experiments on a PC with an Nvidia RTX 2080 graphics card, with an AMD 3950x CPU and 32G memory. Our method takes on average 2.1 ms to synthesize one frame, which is sufficient for real-time applications. As in similar research, real-time in-between motion generation requires high-quality data. Therefore, we mainly use the Lafan1 dataset~\cite{harvey_robust_2020} for its good quality and diversity in motion styles. To further test the generalizability of our method, we also validate it on the Human3.6M dataset~\cite{IonescuSminchisescu11}~\cite{h36m_pami} and compare it with previous methods. Unless specified otherwise, the following experiments are conducted on the Lafan1 dataset and all models are trained with transition lengths of 5 to 30 frames (see supplemental material for details).

Our data split for training/testing is similar to~\cite{harvey_robust_2020}. Each test window contains 65 frames, sampled from Subject 5 of both datasets. Two consecutive windows have 25 overlapped frames. Our evaluation focuses on the motion quality, transition quality and model generalizability under unseen control signals. We employ both qualitative visual evaluation and quantitative metrics. The quantitative metrics include reconstruction accuracy given a seen target frame and transition duration, evaluated by Normalized Power Spectrum Similarity (NPSS) in the joint angle space and the average L2 distance of global joint position between the predicted results and the ground truth. These metrics are good indicators of transition quality, i.e., testing whether multi-modal transitions are captured in detail. {Note that although we add a bone-length loss term during training, it cannot keep the lengths of the bones constant. For a 30-frames transition motion, the average bone length error is $0.64$ cm. For a fair comparison with RTN, we first transform the joint positions to joint rotations, and then obtain their joint positions by FK before we compare their joint position accuracy.} In addition, we also employ a foot skating metric to evaluate the motion quality~\cite{zhang_mode-adaptive_2018}. This metric checks whether the motion manifold learns a reasonable pose under a given velocity. The foot skating metric averages the foot velocity $v_f$ over the ground if the foot height $h$ is within a threshold $H$. Since there is foot skating in the ground-truth, we empirically set $H$ to 2.5 cm. The metric is defined as:
\begin{equation}
    L_{f}=v_f\cdot \operatorname{clamp}(2-2^{h/H},0,1).
\end{equation}

\subsection{Ablation study}

% We first show that the motion manifold can always sample a pose that adapts to the given hip velocity to ensure the motion quality even the transition sampler is not well trained. We validate this by checking the generated sequence of transition sampler at the different iterations. As the number of iterations increases, the previous network\cite{harvey_robust_2020}\cite{harvey2018recurrent} generates the pose, gradually becoming reasonable and approaching the target frame simultaneously.  Our architecture always generate reasonable pose during the training process but there is a gap to the target frame at first. As the number of iterations increases, the sampler learns to approach to target frame.

\textbf{Pose Representation}
Previous research uses joint positions, joint angles or both~\cite{holden_learned_2020,holden2017phase}. To test which representation works the best for our manifold model, we conduct an ablation study and focus on the foot skating, shown in Table \ref{tab:representation}. The results are similar to existing research~\cite{Wang_STRNN_2019} in that joint position representation can effectively mitigate the foot skating. Note that our model still explicitly models joint angles, which is normally required for animation purposes. Using joint positions here acts as a regularization term to facilitate learning. 

\begin{table}[]
\caption{Comparisons of foot skating between position-based representation and rotation-based representation.}
\label{tab:representation}
\tiny
\centering
\resizebox{0.95\columnwidth}{!}{%
\begin{tabular}{llll}
 &\multicolumn{3}{c}{\textbf{Foot skate}} \\ \cline{2-4}
\multicolumn{1}{l}{Frames} &
  \multicolumn{1}{l}{5} &
  \multicolumn{1}{l}{15} &
  \multicolumn{1}{l}{30} \\ \hline
% Interpolation  &  1.709 &  2.080 &  2.144   \\ 
Rotation-based  &  0.934 &  1.035 &  1.161   \\ 
Position-based &  \textbf{0.356} &  \textbf{0.373} &  \textbf{0.401 } 
\end{tabular}}
\end{table}

\textbf{Motion manifold focus on the lower-body joints.}
Our focus on the lower-body and the hip in natural motion modeling is a different design choice compared with recent deep learning research. This is because the lower-body motions are relatively simple in locomotion but highly important for motion quality due to foot skating. Therefore, we prioritize these joints in learning the manifold and leave the correlation learning between the upper-body and the lower-body to the transition sampler. Alternatively, we can also model the whole body directly but the relatively unconstrained upper-body motion can introduce ambiguity in learning. 

To prove this, we add the upper-body joints to the CVAE and remove the upper joints' rotation difference $\dot{\bold{r}}_U$ from the output of the transition sampler. We refer to this network as the Full-body network. A comparison is shown in Table~\ref{tab:divide}. Smaller reconstruction errors in both the joint angle and position space indicate that better transitions are learned. Note that in both networks, our method still predicts the full body. The larger errors in Full-body are likely to be caused by its predictions being closer to the `averaged' motion. By separating the modeling of the lower-body and upper-body joints, our method manages to improve the learning.

\begin{table}[tb]
\caption{Comparisons between the Full-body network and our method. Both models are trained with 300,000 iterations.}
\label{tab:divide}
\tiny
\centering
\resizebox{0.95\columnwidth}{!}{%
\begin{tabular}{llll}
& \multicolumn{3}{c}{\centering\textbf{L2 norm of global position}} \\ \cline{2-4}
\specialrule{0em}{1pt}{0pt} \multicolumn{1}{l}{Frames} &
  \multicolumn{1}{l}{5} &
  \multicolumn{1}{l}{15} &
  \multicolumn{1}{l}{30} \\ \hline
{Full-body} &  0.259 &  0.612 &  1.143  \\
Our method & \textbf{0.196} &  \textbf{0.562} &  \textbf{1.124}  
\\ \hspace*{\fill} 

& \multicolumn{3}{c}{\centering\textbf{NPSS}} \\ \cline{2-4}

\specialrule{0em}{1pt}{0pt}
{Full-body}      &  0.00671 & 0.07471 & 0.35385 \\ 
Our method & \textbf{0.00554} & \textbf{0.07026} & \textbf{0.34549}  
%  \\ \hspace*{\fill} 
 
% & \multicolumn{3}{c}{\centering\textbf{Foot skate}} \\  \cline{2-4}
% \specialrule{0em}{1pt}{0pt} Ground Truth &  0.162 &  0.141 &  0.143 \\ 
% {Full-body}      &  \textbf{0.226} & \textbf{0.318} & \textbf{0.465} \\ 
% Our method &  0.244(-0.02) &  0.343(-0.03) &  0.469(-0.004) \\ 
\end{tabular}}
\vspace{-2.0em}
\end{table}

\subsection{Evaluation and Comparison}

% We linearly interpolate the start frame and the target frame as the baseline.

\begin{table}[]
\tiny
\centering
\caption{Comparisons of reconstruction accuracy and foot skating of different methods. All models are trained with 300,000 iterations.}
\label{tab:lengths}

\resizebox{0.95\columnwidth}{!}{%
\begin{tabular}{llll}
& \multicolumn{3}{c}{\centering\textbf{L2 norm of global position}} \\ \cline{2-4}
\specialrule{0em}{1pt}{0pt} \multicolumn{1}{l}{Frames} &
  \multicolumn{1}{l}{5} &
  \multicolumn{1}{l}{15} &
  \multicolumn{1}{l}{30} \\ \hline
% {Interpolation}  & 0.371  &1.243  & 2.306    \\ 
% {RTN}  &  0.215 &  0.591 &  1.163  \\
% {+skating loss}  &  0.279 &  0.683 &  1.266  \\ 
% {Auto-Encoder} &  0.278 &  0.625 &  1.156   \\ 
%  {VAE} &  \textbf{0.195} &  \textbf{0.556} &  \textbf{1.110}\\
% Our method & {0.196} &  {0.562} &  {1.124}  
% \\ \hspace*{\fill} \\

% & \multicolumn{3}{c}{\centering\textbf{NPSS}} \\ \cline{2-4}

% \specialrule{0em}{1pt}{0pt}{Interpolation} & 0.00728 & 0.11350 & 0.52291  \\ 
% {RTN} & 0.00559 & 0.07194 & 0.34952   \\ 
% {+skating loss} & 0.00706 & 0.07993 & 0.37151  \\
% {Auto-Encoder} & 0.00784 & 0.08345 & 0.37210 \\ 
% {VAE} & \textbf{0.00554} & \textbf{0.07005} & \textbf{0.34199}\\
% Our method & \textbf{0.00554} & {0.07026} & {0.34549}  \\ 

% \\
% & \multicolumn{3}{c}{\centering\textbf{Foot skate}} \\  \cline{2-4}
% \specialrule{0em}{1pt}{0pt} Ground Truth &  0.162 &  0.141 &  0.143 \\ 
% {Interpolation}  &  1.708 &  2.081 &  2.144  \\ 
% {RTN}   &  0.483 &  0.698 &  0.930  \\ 
% {+skating loss} &  0.249 &  0.349 &  \textbf{0.455} \\ 
% {Auto-Encoder} &  0.294 &  0.485 &  0.649 \\ 
%  {VAE} &  0.255 &  0.353 &  0.502\\
% Our method &  \textbf{0.244} &  \textbf{0.343} &  0.469 \\ 

{Interpolation}  & 0.37  &1.24  & 2.31    \\ 
{RTN}  &  0.22 &  0.59 &  1.16  \\
{+skating loss}  &  0.28 &  0.68 &  1.27  \\ 
{Auto-Encoder} &  0.28 &  0.63 &  1.16   \\ 
 {VAE} &  \textbf{0.20} &  \textbf{0.56} &  \textbf{1.11}\\
Our method & \textbf{0.20} &  \textbf{0.56} &  {1.12}  
\\ \hspace*{\fill} \\

& \multicolumn{3}{c}{\centering\textbf{NPSS}} \\ \cline{2-4}

\specialrule{0em}{1pt}{0pt}{Interpolation} & 0.0073 & 0.1135 & 0.5229  \\ 
{RTN} & 0.0056 & 0.0719 & 0.3495   \\ 
{+skating loss} & 0.0071 & 0.0799 & 0.3715  \\
{Auto-Encoder} & 0.0078 & 0.0835 & 0.3721 \\ 
{VAE} & \textbf{0.0055} & \textbf{0.0701} & \textbf{0.3420}\\
Our method & \textbf{0.0055} & {0.0702} & {0.3455}  \\ 

\\
& \multicolumn{3}{c}{\centering\textbf{Foot skate}} \\  \cline{2-4}
\specialrule{0em}{1pt}{0pt} Ground Truth &  0.162 &  0.141 &  0.143 \\ 
{Interpolation}  &  1.708 &  2.081 &  2.144  \\ 
{RTN}   &  0.483 &  0.698 &  0.930  \\ 
{+skating loss} &  0.249 &  0.349 &  \textbf{0.455} \\ 
{Auto-Encoder} &  0.294 &  0.485 &  0.649 \\ 
 {VAE} &  0.255 &  0.353 &  0.502\\
Our method &  \textbf{0.244} &  \textbf{0.343} &  0.469 \\

\end{tabular}}
\vspace{-2.0em}
\end{table}

\subsubsection{Motion quality.}
Foot sliding is an important metric for motion quality. Earlier research suffers from such problems due to `averaged' motions and drifting issues~\cite{fragkiadaki2015recurrent}. A common strategy is to have a post-processing stage~\cite{Wang_STRNN_2019}. More recent work tends to mitigate this problem e.g. by inducing contact patterns~\cite{starke_local_2020} when dense temporal control signal is available, or constraining the distribution of the generated motions to be similar to that of the data~\cite{harvey_robust_2020}. However, while post-processing is highly undesirable in our application, predicting contact patterns is also not straightforward with arbitrary target frames. Constraining the distribution seems effective, but it is still not easy to mitigate the foot skating.
To show this, we compare RTN~\cite{harvey_robust_2020} and its variant with a foot skating loss (+skating loss). As a naive baseline, we also experiment with linear interpolation (Interpolation). Results are shown in Table~\ref{tab:lengths}. First, adding an additional foot skating loss to RTN mitigates foot skating to some extent (Table~\ref{tab:lengths} Bottom). However, it also leads to worse reconstruction accuracy. During learning, the predicted contact step might be different from the step from the ground truth, so their loss induces an incorrect backward gradient and make the pose unnatural. The results in the video show that adding the foot skating loss also sometimes causes unsmoothed transitions near the start and the end frame. Although our method predicts the foot contact step, the inconsistent backward gradient doesn't affect the pose unnaturally but helps to adjust the hip velocity because the pose is sampled from the manifold. 

Further, to also show the importance of controlling the latent distribution of the motion dynamics, we replace our CVAE with a plain autoencoder (Auto-Encoder) so that the distribution of $z$ is not constrained (i.e., without the KL-divergence loss). Besides, to show the importance of using the hip velocity as a condition, we replace our CVAE with a VAE so that the latent distribution is unconditional. Autoencoder and VAE are widely used for learning motion manifolds~\cite{holden_learned_2020,harvey_robust_2020,Wang_STRNN_2019,ling_character_2020}. The results are shown in Table~\ref{tab:lengths}. By explicitly controlling $z$ with conditioned decoding on the hip velocity, our method performs similarly to the VAE in reconstruction accuracy (i.e., L2 error and NPSS), significantly better than the Auto-encoder. Further, our method outperforms VAE in foot skating by large margins especially from the 30th frame, greatly improving the motion quality. 
% By explicitly controlling $z$ with conditioned decoding on the hip velocity, our method achieves overall better results in both avoiding foot skating and reconstruction accuracy.

% We also compare our method with RTN on the Human3.6M dataset, and compare our method with the state-of-the-art offline motion completion method which is based on a transformer~\cite{duan_single-shot_2021}. The above experiments' results are reported in the supplemental material.

We compare our method with RTN on the Human3.6M dataset, shown in Table~\ref{tab:hm3.6}. While RTN achieves slightly better results in NPSS by as large as 7.6\%, our method outperforms RTN in both the $L_2$ norm and the foot skate by 10.3\% and 69.2\%, respectively. When looking at the difference between the Human3.6M and the Lafan1, the most significant difference is the skeleton variation between subjects. The Human3.6M has a larger variation. We speculate this might be the reason for the slightly worse results in NPSS and also why at frame 5, Interpolation even outperforms both methods. However, since neither methods explicitly aim to generalize to different skeletons, we leave the analysis to future work.

\begin{table}[ht]
\tiny
\centering
\caption{[Human3.6M]Comparisons of different frames between RTN and our method.  Both models are trained with 300,000 iterations.}
\label{tab:hm3.6}

\resizebox{0.95\columnwidth}{!}{%
\begin{tabular}{llll}

& \multicolumn{3}{c}{\centering\textbf{L2 norm of global position}} \\ \cline{2-4}
\specialrule{0em}{1pt}{0pt} \multicolumn{1}{l}{Frames} &
  \multicolumn{1}{l}{5} &
  \multicolumn{1}{l}{15} &
  \multicolumn{1}{l}{30} \\ \hline
{Interpolation} &  0.78 &  1.85 &  2.59   \\ 
{RTN}  &  0.53 &  {0.98} &  {1.50}  \\
Our method &  \textbf{0.47} &  \textbf{0.93} &  \textbf{1.44}
\\ \hspace*{\fill} 

& \multicolumn{3}{c}{\centering\textbf{NPSS}} \\ \cline{2-4}

\specialrule{0em}{1pt}{0pt}
{Interpolation}  & \textbf{0.0044} & 0.0651 & 0.3132  \\ 
{RTN} & {0.0049} & \textbf{0.0549} & \textbf{0.2298}  \\ 
Our method  & 0.0054 & 0.0554 & 0.2386  \\ 

& \multicolumn{3}{c}{\centering\textbf{Foot skate}} \\  \cline{2-4}
\specialrule{0em}{1pt}{0pt} 
Ground Truth &  0.070 &  {0.089} &  {0.089} \\ 
{Interpolation}  &  0.824 &  1.525 &  0.929 \\ 
{RTN}  &  0.325 &  0.363 &  0.456  \\ 
Our method &  \textbf{0.100} &  \textbf{0.198} &  \textbf{0.317} \\ 

\end{tabular}}
\end{table}

In addition, we also compare our method with the state-of-the-art offline motion completion method, which is based on a transformer~\cite{duan_single-shot_2021}. The results reported in Table~\ref{tab:transformer} shows that our method can generate comparable results. 
% \HW{The reviewer could raise questions about this comparison as we only used the $L_2$ metric, inconsistent with other experiments.}

\begin{table}[ht]

\tiny
\centering
\caption{Comparisons between our method and the transformer. }
\label{tab:transformer}

\resizebox{0.95\columnwidth}{!}{%
\begin{tabular}{llll}
& \multicolumn{3}{c}{\textbf{L2 norm of global position}} \\ \cline{2-4}
\specialrule{0em}{1pt}{0pt}\multicolumn{1}{l}{Frames} &
  \multicolumn{1}{l}{5} &
  \multicolumn{1}{l}{15} &
  \multicolumn{1}{l}{30}\\ \hline  
\specialrule{0em}{1pt}{0pt}Transformer  & 0.22 &  \textbf{0.56} &    \textbf{1.10}  \\ 
Our method &\textbf{0.20}&  \textbf{0.56} &  1.12  \\ 
\end{tabular}}
\vspace{-2.0em}
\end{table}

\subsubsection{Generalization}
Since the target frame and aimed transition duration can be used as control signals, evaluating models under unseen and extreme user control is crucial. We validate our method via multiple aspects: motion style, aimed transition duration and the distance between the start and end frame.

\begin{table}[t]
\tiny
\centering
\caption{Comparisons of different kinds of actions.
 The number of frames of the generated sequences is 30. Both models are trained with 300,000 iterations.\label{tab:actions}}

\resizebox{1.0\columnwidth}{!}{%
\begin{tabular}{lllll}
 &\multicolumn{4}{c}{\textbf{L2 norm of global position}} \\ \cline{2-5}
\specialrule{0em}{1pt}{0pt}\multicolumn{1}{l}{Actions} &
  \multicolumn{1}{l}{Walk} &
  \multicolumn{1}{l}{Dance} &
  \multicolumn{1}{l}{Jump} &
  \multicolumn{1}{l}{Obstacle} \\ \hline
% Interpolation  &  2.760  &  2.391 &  1.887&  2.233 \\ 
% RTN  &  0.992 &  1.509  &  1.212  &  1.212 \\ 
% Our method   &  \textbf{0.948}  &  \textbf{1.477}  &  \textbf{1.183}  &  \textbf{1.150}  \\ 

%  &\multicolumn{4}{c}{\centering\textbf{NPSS}} \\ \cline{2-5}
% \specialrule{0em}{1pt}{0pt}Interpolation  & 0.64304 & 0.64050 & 0.39995  & 0.45134  \\ 
% RTN & 0.33799  & 0.51970 & \textbf{0.31229}    & 0.31577  \\ 
% Our method & \textbf{0.33064} & \textbf{0.51409}  & 0.32052   & \textbf{0.30853} \\ 

%   &\multicolumn{4}{c}{\centering\textbf{Foot skate}} \\ \cline{2-5}
% \specialrule{0em}{1pt}{0pt} Ground Truth&  0.160 &  0.230 &  0.155   &  0.121  \\
% Interpolation   &  2.743  &  1.844 &  1.381 &  2.024  \\ 
% RTN  &  1.187 &  1.103   &  0.640   &  0.657  \\ 
% Our method  &  \textbf{0.589}  &  \textbf{0.571}   &  \textbf{0.326}  &  \textbf{0.293}  \\ 

Interpolation  &  2.76  &  2.40 &  1.89&  2.23 \\ 
RTN  &  0.99 &  1.51  &  1.21  &  1.21 \\ 
Our method   &  \textbf{0.95}  &  \textbf{1.48}  &  \textbf{1.18}  &  \textbf{1.15}  \\ 

 &\multicolumn{4}{c}{\centering\textbf{NPSS}} \\ \cline{2-5}
\specialrule{0em}{1pt}{0pt}Interpolation  & 0.6430 & 0.6405 & 0.4000  & 0.4513  \\ 
RTN & 0.3380  & 0.5197 & \textbf{0.3123}    & 0.3158  \\ 
Our method & \textbf{0.3306} & \textbf{0.5141}  & 0.3205   & \textbf{0.3085} \\ 

  &\multicolumn{4}{c}{\centering\textbf{Foot skate}} \\ \cline{2-5}
\specialrule{0em}{1pt}{0pt} Ground Truth&  0.160 &  0.230 &  0.155   &  0.121  \\
Interpolation   &  2.743  &  1.844 &  1.381 &  2.024  \\ 
RTN  &  1.187 &  1.103   &  0.640   &  0.657  \\ 
Our method  &  \textbf{0.589}  &  \textbf{0.571}   &  \textbf{0.326}  &  \textbf{0.293}  \\ 

\end{tabular}}
\end{table}

\textbf{Different motion styles.}
We split the Lafan1 dataset into multiple subsets by motion styles. All the walking and running sequences are categorized into a 'walk' set. The 'dance' set includes dance sequences, fight sequences and sports sequences. A 'jump' set collects all the jump sequences, and an 'obstacles' set contains all the character's motions across the obstacles. The results are shown in Table~\ref{tab:actions}. Since the ground truth data also contains foot skating, we use it as a baseline too. The results show that our method can universally improve the results. While being able to achieve better reconstruction, our method provides far better foot skates. This is demonstrated across different motion styles.

\textbf{Different transition duration}. To test our method can handle long and short duration as the aimed transition time, we discard 1 second (30 frames) from each sample, and require the network to generate the sequences at speed scaled to 2x (15 frames), 4x (8 frames) and 0.5x (60 frames) respectively. Besides, we have made experiments with an extreme case, in which we slow down the speed 100 times (0.01x). The results are shown in Table~\ref{tab:changeTime}. Interpolation performs best in this situation because the drifting is divided by 100 times. However, there is no valid pose change in interpolation results.

Slowing down motions is theoretically easier as foot contact can still be generated. Our method outperforms RTN in keeping foot contact. Comparatively, speeding up is more difficult because it might be impossible for a character to reach the target during such a short time. This is why the foot skating in 4x and 2x are in general worse. However, our method still provides the best results. Visually, we find that if the start and the end frame are within the same phase of a walking cycle, for example, when the left leg is behind the right leg, RTN motion tends to drift to the target without changing pose, while our results make fast footsteps. These results can be seen in the accompanying video.

\begin{table}[tb]
\tiny
\centering
\caption{Comparisons of foot skating of changing transition duration between RTN and our method. Both models are trained with 300,000 iterations.}
\label{tab:changeTime}

\resizebox{1.0\columnwidth}{!}{%
\begin{tabular}{lllll}
& \multicolumn{4}{c}{\textbf{Foot skate}} \\ \cline{2-5}
\specialrule{0em}{1pt}{0pt}\multicolumn{1}{l}{Speed} &
  \multicolumn{1}{l}{4x} &
  \multicolumn{1}{l}{2x} &
  \multicolumn{1}{l}{0.5x}&
   \multicolumn{1}{l}{0.01x}\\ \hline  
Ground Truth &0.161&0.148&0.156& 0.149\\ 
Interpolation &7.302&3.917&1.075&\textbf{0.004}\\ 
RTN  & 4.050 &  2.087 &    0.814 & 0.522 \\ 
Our method &  \textbf{3.363} &  \textbf{1.350} &    \textbf{0.438} & {0.110} \\ 
\end{tabular}}
\vspace{-2.0em}
\end{table}

\textbf{Different target locations}. Another dimension is to vary the location of the target frame. We conduct two experiments. The first one has the target frame set further away, along the direction from the start to the target frame (Forwarding). In addition, we also move the target frame in the opposite direction (i.e., from the original target frame to the starting frame), to a location that is twice further away from the starting frame compared with the original target frame (Backwarding).

In Forwarding, our method generates fewer bigger steps or more small steps to fill the gap without visible artifacts. RTN usually performs at the same pace as the ground truth but fills the distance gap by drifting. Backwarding is an extreme testing case and is challenging for both methods. Since the poses of the start frame and the target frame have similar orientations and there is not enough time for turning twice, both methods try to achieve the target frame by walking backward. However, RTN always generates visible artifacts while our method does not. We speculate that this is because the data does not contain enough clips for the character to move backwards. However, the motion manifold captured the motions with backward velocity in the training set, which helps to synthesize the natural action successfully. 
Results of both Forwarding and Backwarding can be found in the video. Here, we show an extreme Forwarding case where the target frame is $10$ meters away from the starting frame, where the longest distance in the training set is merely $5.79$ meters. To generate a 60-frames motion, the character must have a speed exceeding $5m/s$ ($7.9m/s$ for the farthest distances) to reach the target. 

\begin{figure}[t]
	\centering
	\includegraphics[width=\linewidth]{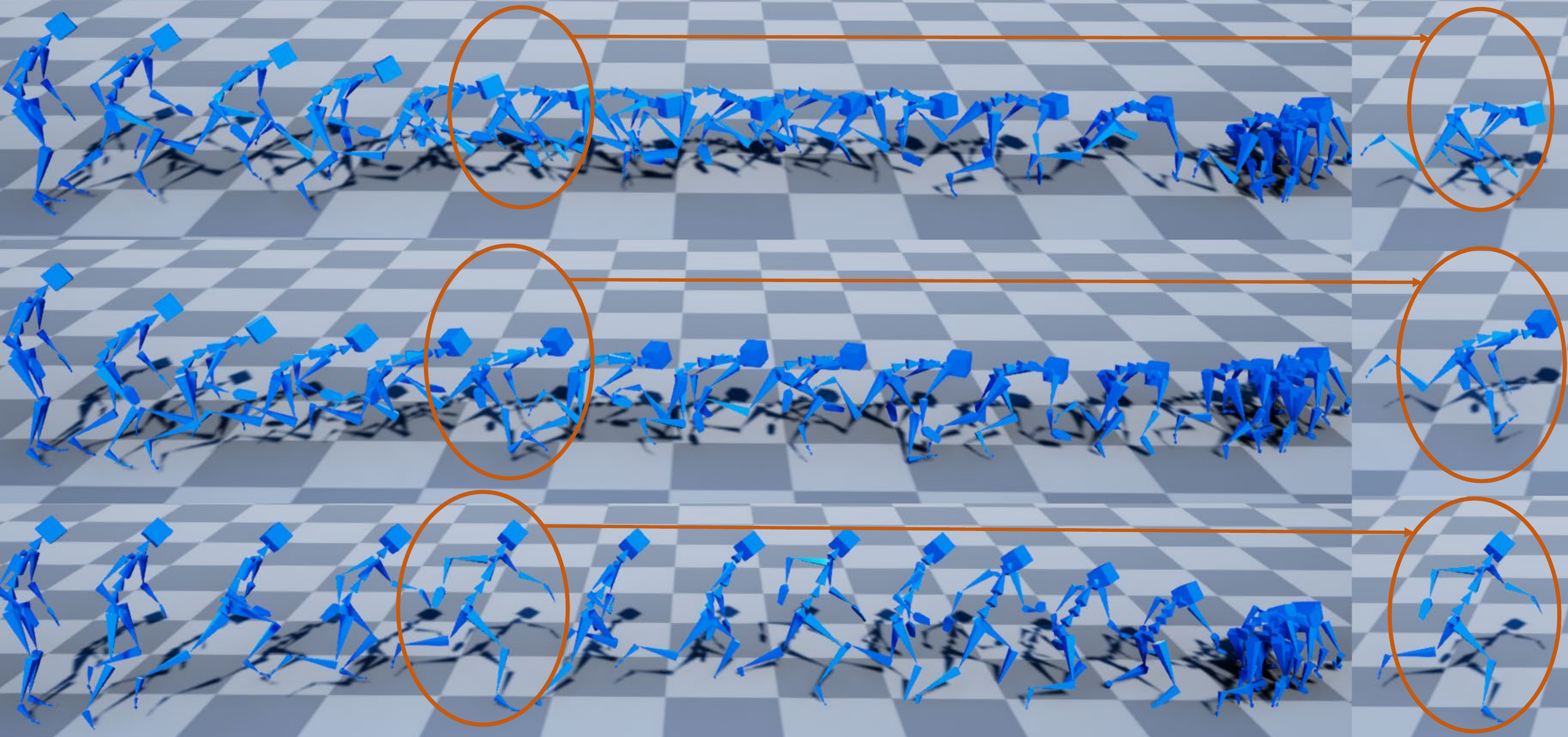}
    \caption{We sample the pose every 5 frames. The first row shows the RTN's results, the second row shows the VAE's results, and the last row shows our results. For a clearer observation, we sample one frame from each sequence and put it at the rightmost row. The results indicate that our method generates the most natural running motion. }
	\label{fig:longdistance}
\end{figure}

Figure~\ref{fig:longdistance} shows a visual comparison among RTN, VAE and our method. RTN (the first row) generates motions that drift towards the goal because the target frame is too far (i.e., floating in the air in the middle of the motion). In addition, without the condition of the hip velocity, the poses synthesized by VAE do not conform to the movements, leading to unnatural motions (i.e., the poses are similar to the blending of crawling and running). 
% Without the condition of the hip velocity, the VAE baseline, more precisely the transition sampler in this baseline, is not guided, leading to unnatural motions such as a sudden acceleration (i.e. the big gap between the third and the fourth frame). In addition, RTN generates motions that drift towards the goal because the target frame it too far (i.e. floating in the air in the middle of the motion).
Comparatively, our method generates the most natural motions under this extreme case.

% In addition, to show that encoding the pose distribution under hip velocity improves the generalization in the wild, we replace our CVAE with a VAE by discarding the hip velocity to the decoder.  However, it cannot generate a natural pose under the "strange" hip velocity, which is different from the velocity distribution of the dataset, as shown in Figure~\ref{fig:longdistance}.

\section{Limitations}

Similar to other data-driven methods, one major limitation of our method is that our generation results are limited by the training data. It cannot generate motions that are too different from the training data. One example is that motion generation will become quite challenging when we place the target frame behind the starting frame. The reason is that there is no quick turning or enough backward walking motion in the training data. Another example is that we cannot guarantee the target frame is 100\% achieved if the target frame is too different from data, spatially, temporally, or both. However, we argue that our framework itself is still effective and the aforementioned problems can be easily overcome when more diverse data are introduced.

\begin{figure}[t]
	\centering
	\includegraphics[width=0.9\linewidth]{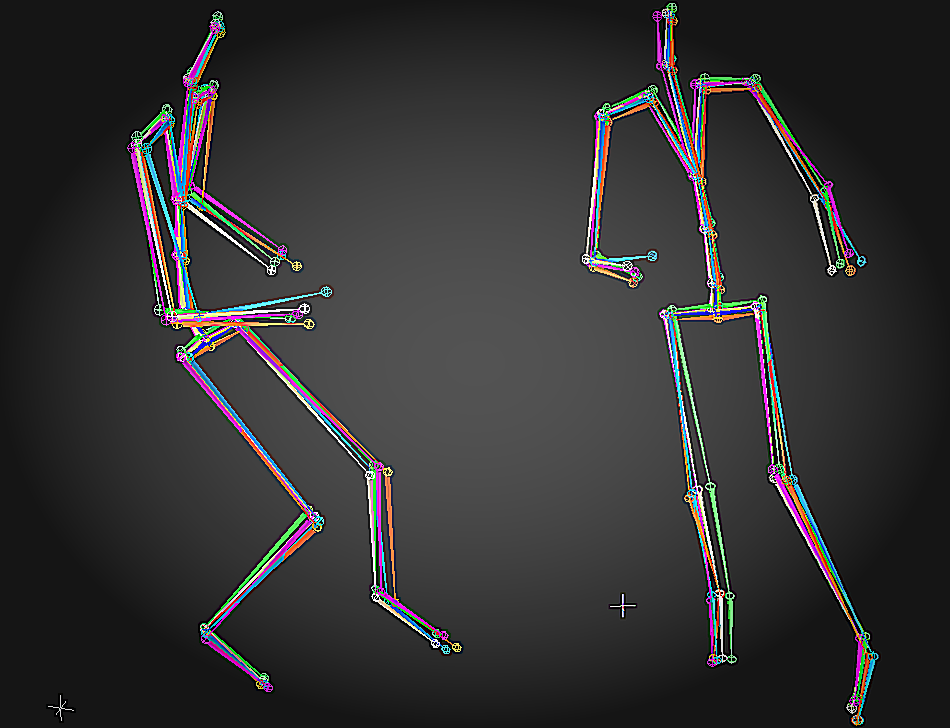}
    \caption{The figure shows the midway in a 30-frames transition re-sampled 5 times. The upper body shows the diversity. }
	\label{fig:diversity}
	\vspace{-1.0em}
\end{figure}

{Lacking of motion diversity is another limitation of our method. As a CVAE-based network, our model can indeed generate different motions for the same control, but the differences of the generated motions are small, especially for the lower body (as shown in Figure~\ref{fig:diversity}). To generate high-quality motion under unseen control, we assume that the contact position with the ground of each step cannot change too much.}

\section{Future work}
{Although specifying the motion duration is widely used in the animation/game pipelines to control the timing of motions and transitions, we will provide automated computation for desired motion duration in future. Given the robustness of our model under different timing requirements, we will compute a reasonable duration by modeling the distribution of possible timing requirements automatically. In addition, we will take more factors into considerations, such as motion styles, interactions with environments, and various skeletal topologies (e.g. quadrupeds).}

\section{Conclusions}
We proposed a novel learning framework consisting of a new natural motion manifold model and a new transition sampler for real-time in-between motion generation. The motion manifold model treats different body parts separately and focuses on controllability and motion quality, while the transition sampler ensures natural motions are generated with respect to user control. Our model generates high-quality motions in mitigating foot skating and motion transitions {so that it can be used for both offline animation and online games.} Our method is also general under unseen control signals. It outperforms alternative solutions and the state-of-the-art methods.

\begin{acks}
Xiaogang Jin was supported by  the National Natural Science Foundation of China (Grant Nos. 62036010, 61972344), the Ningbo Major Special Projects of the “Science and Technology Innovation 2025” (Grant No. 2020Z007), and the Key Research and Development Program of Zhejiang Province (Grant No.   2020C03096).
\end{acks}
\bibliographystyle{ACM-Reference-Format}
\bibliography{sample-base}

\appendix

\end{document}